
%
%
%
\font\subtit=cmr12
\font\name=cmr8
\input harvmac
\def\MPILMU#1#2#3#4
{\TITLE{MPI-Ph/\number\yearltd-#1}
{LMU-TPW \number\yearltd-#2}{#3}{#4}}
\def\TITLE#1#2#3#4{\nopagenumbers\abstractfont\hsize=\hstitle\rightline{#1}
\vskip 1pt\rightline{#2}
\vskip 1in
\centerline{\subtit #3}
\vskip 1pt
\centerline{\subtit #4}\abstractfont\vskip .5in\pageno=0}%
\MPILMU{50}{14}
{ON THE QUANTIZATION OF ABELIAN GAUGE FIELD}{THEORIES ON RIEMANN SURFACES}

\centerline{F{\name RANCO} F{\name ERRARI}\foot{\name This work
is carried out in the framework of the EC Research Programme
``Gauge theories, applied supersymmetry and quantum gravity".}}\smallskip
\centerline{\it Max-Planck-Institut f\"ur Physik}\smallskip
\centerline{\it - Werner-Heisenberg Institut - }\smallskip
\centerline{\it P.O.Box 401212, Munich (Fed. Rep. Germany)}
\vskip 1cm
\centerline{ABSTRACT}
{\narrower
In this paper we quantize the abelian gauge field theories on a Riemann
surface $M$ in the Feynman gauge. The fields take their values on any
nontrivial line bundle $P(M,U(1))$ with first Chern class $c_1=2\pi k$,
$k\in {\rm\bf Z}$. The point of view adopted here is that of small
quantum perturbations $A^{\rm qu}$ around a classical instantonic
solution $A^{\rm cl}\in P(M,U(1))$.
The explicit form of the instantonic fields $A^{\rm cl}$ and of the
propagator $\langle A^{\rm qu} A^{\rm qu}\rangle$ is derived.
The case in which the theory interacts only with an external current $J$
is completely solved evaluating the generating functional $Z[J]$.
Finally, we consider the Schwinger model, or two dimensional quantum
electrodynamics. We show that it is possible to integrate out from the
path integral the
nonphysical gauge degrees of freedom . The upshot is a nonlocal field
theory of fermions describing the dynamics of the electrons on a Riemann
surface.
The fermions interact through a potential whose short distance behavior
is explicitly derived.
In particular, it turns out that this behavior can only depend on
zero modes and in the case of the sphere, where there are not zero modes,
the interactions
between the electrons switch off at very high energies yielding a free
field theory.
}
\Date{July 1993}
\newsec { INTRODUCTION}
\vskip 1cm
One of the main motivations for studying the gauge field
theories in two dimensions
is that they provide a good laboratory in order to illustrate
some important
properties
which are present also in the more
realistic four dimensional models,
like anomalies, confinement and dynamical symmetry breaking
\ref\flat{S. Coleman, {\it Phys. Rev.} {\bf D11} (1975), 3026;
S. Donaldson, {\it J. Diff. Geom.}
{\bf 18} (1983), 269;
R. Jackiw, R. Rajaraman, {\it Phys. Rev. Lett.} {\bf 54} (1985), 1219;
L. Faddev, S. Shatashvili, {\it Phys. Lett.} {\bf 183B} (1987), 311;
T. P. Killingback, {\it Phys. Lett.} {\bf 223B} (1989) 357;
A. I. Bocharek, M. E. Schaposhnik, {\it Mod. Phys. Lett.} {\bf A2} (1987),
991; J. Kripfganz, A. Ringwald, {\it Mod. Phys. Lett.} {\bf A5}
(1990), 675.}.
Moreover, following the recent developments in string theories
and
topological field theories, there is also an interest in
quantizing the two dimensional field theories on a manifold and in particular
on the closed and orientable Riemann surfaces
\ref\witt{E. Witten, {\it Comm. Math.
Phys.} {\bf 141} (1991), 153; {\it Two
Dimensional Gauge Theories Revisited}, Preprint IASSNS-Hep
92/15.},
\ref\rusakov{
D. J. Gross, W. Taylor IV, Twist and Wilson Loops
in the String Theory of Two Dimensional QCD, Preprint CERN-TH 6827/93,
PUPT-1382, LBL-33767, UCB-PTH-93/09, March 1993;
B. Ye. Rusakov, {\it Mod. Phys. Lett.} {\bf A5}
(1990), 693; A. Migdal, {\it Zh. Esp. Teor. Fiz.} {\bf 69} (1975),
810.}, \ref\froe{J. Fr\"ohlich, On the Construction of Quantized
gauge Fields, in ``Field Theoretical Methods in Particle Physics" (W.
R\"uhl Ed.) Plenum, New York, 1980;
I. L. Buchbinder, S. D. Odintsov, I. L. Shapiro,
{\it Effective Action in Quantum Gravity}, IOP Publishing, Bristol and
Philadelphia, 1992.},
\ref\bt{M. Blau, G. Thompson, {\it Int. Jour. Mod. Phys.}
{\bf A7} (1992), 3781.}, \ref\thom{G. Thompson, 1992 Trieste Lectures on
Topological Gauge Theory and Yang-Mills Theory.},
\ref\fine{D. S. Fine, {\it Comm. Math. Phys.} {\bf 134} (1990), 273;
S. G. Rajeev, {\it Phys. Lett.} {\bf 212B} (1988), 203.},
\ref\fp{
D. Z. Freedman, K. Pilch, {\it Phys. Lett.}{\bf 213B} (1988), 331.},
\ref\tomboulis{E. T. Tomboulis, {\it Phys. Lett.} {\bf 198B} (1987),
165; M. Porrati, E. T. Tomboulis, {\it Nucl. Phys.} {\bf B315} (1989), 615.}.
Unfortunately, most of the literature in string theory deals with the
pure Yang-Mills field theory, in which the observables
are metric independent objects like the Wilson loops and, more in general,
our knowledge about
the interacting case is confined until now
to very simple topologies.
For example the Schwinger model or two dimensional quantum
electrodynamics ${\rm QED}_2$ has been understood only in the flat
case
\ref\cmpl{J. H. Lowenstein, J. A. Swieca, {\it Ann. Phys.} {\bf 68}
(1961), 172;
A. Z. Capri, R. Ferrari, {\it Nuovo Cim.} {\bf 62A} (1981), 273;
{\it Journ. Math. Phys.} {\bf 25} (1983), 141;
G. Morchio, D. Pierotti, F. Strocchi, {\it Ann. Phys.} {\bf 188}
(1988), 217;
A. K. Raina, G. Wanders,
{\it Ann. of Phys.} {\bf 132} (1981), 404.}, on the cylinder
\ref\manton{N. Manton, {\it Ann. Phys.} {\bf 159} (1985), 220; J. E.
Hetrick, Y. Hosotani, {\it Phys. Rev.} {\bf 38} (1988), 2621.},
on the sphere
\ref\jaw{C. Jayewardena, {\it Helv. Phys. Acta} {\bf 61} (1988), 636.}
and on the torus
\ref\Joos{H. Joos, {\it Nucl. phys.} {\bf B17}
(Proc. Suppl.) (1990), 704; {\it Helv. Phys. Acta} {\bf 63} (1990),
670},
\ref\wipf{I. Sachs, A. Wipf, {\it Helv. Phys. Acta} {\bf 65} (1992),
653.}.\smallskip
To overcome at least in part these limitations,
we propose a perturbative approach
which is able to quantize the interacting abelian gauge fields
on a Riemann surface, even in the presence of
nontrivial topological sectors.
This approach, which is the main result of
this paper, allows for example the quantization of
models containing massless scalar field theories and fermions. The
quantization of other abelian gauge field theories, in which for instance
the matter fields are massive, is however
restricted by the fact that the
propagators of these fields are not known on a Riemann surface.
Another important feature of the formalism presented here,
is that it is very explicit.
For this reason,
also phenomenological
considerations concerning the effects
of a nontrivial
gravitational background on field theories
are possible
(on this topic see as an introduction
refs.
\ref\bd{see for example on this subject the Introduction of
N. D. Birrel, P. C. W. Davies, Quantum
Fields in Curved Space, Cambridge University Press, Cambridge (1982); S.
A. Fulling, Aspects of Quantum Field Theory in Curved Space-Time,
Cambridge University Press, Cambridge (1989).}).
The search of these effects is in fact the second aim of this paper.
To this purpose we study the case of the chiral
Schwinger model
\ref\sch{J. Schwinger,
{\it Phys. Rev.} {\bf 128} (1962), 2425.},
comparing the high energy
behaviors of the chiral fermions on a Riemann surface with the analogous
results
obtained considering other topologies.
\smallskip
The approach followed here to quantize the abelian gauge field theories
is a generalization of
\ref\ferqed{F. Ferrari, {\it Class. Q. Grav.},
{\bf 10} (1993), 1065.}, where the propagators and the vertices of
the Schwinger model
in the Lorentz gauge were firstly computed on a Riemann surface
giving a way of deriving also
the higher order contributions to the correlation functions.
Despite of this success, there are in ref. \ferqed\ some problems
in treating models which contain scalar fields and, more important,
the case of nontrivial
topological sectors was not considered.
To overcome these difficulties, we exploit here the Feynman gauge
instead of the
Lorentz gauge. It is important to stress at this point that
the Lorentz and Feynman gauges are probably
the only gauges in which the equations of motion satisfied by the
propagator of the gauge fields become simple enough to be solved.
The disadvantage of the Lorentz gauge is however the fact that
the propagator acts only
in the space of the transverse degrees of freedom. For this reason one has
always to separate the transverse and longitudinal components of the
gauge fields
operating a different perturbative treatment for both of them. This is
an unnecessary complication for example when the matter
fields are massless scalars. In this case, in fact, the unphysical
longitudinal components of the gauge fields remain coupled to the matter
fields at any perturbative order and it is difficult to show that they do
not contribute to the physical amplitudes.
Fortunately, this problem is not present in the Feynman gauge
where
the propagator is orthogonal only with respect
to the harmonic part of the gauge fields. The zero modes are also
unphysical but they represent only a discrete number of degrees of
freedom which is easy to cope with.\smallskip
Another progress made with respect to ref. \ferqed\ is
that we consider in this work also gauge fields belonging to
nontrivial line bundles. These line bundles are
characterized by a nonvanishing value of the first Chern class $c_1$.
Of course, nontrivial line bundles with $c_1\ne 0$
are not visible within the frame of perturbation theory.
Nevertheless we are still allowed to
consider small
perturbations $A_\alpha^{\rm qu}$ around a classical gauge field
$A^{\rm cl}$ with $c_1\ne 0$.
This is the point of view adopted here. Despite of the fact that
the fields $A^{\rm cl}$
cannot be globally
defined on $M$, we have not found any problem in fixing the Lorentz
gauge
$\partial^\mu A_\mu=0$ in a
global way,
basically because this
gauge fixing does not depend on the fields themselves
but on their derivatives.
Moreover, the instantonic solutions $A^{\rm cl}$ which we explicitly
construct here, satisfy exactly the two dimensional Maxwell equations in
the Feynman gauge.
It is remarkable that a classical field $A^{\rm cl}_\mu$
of this kind decouples from the kinetic part of the action and
appears only in the interaction with the matter fields.
This fact allows the computation of
the propagator of the gauge fields $A^{\rm qu}$
as in the case of the trivial topological sector and, consequently,
the perturbative approach explained in ref. \ferqed\ becomes realizable
also for nontrivial line
bundles.
\smallskip
After these improvements with respect to ref. \ferqed,
it is possible to study the effects of the gravitational background
provided by a Riemann surface on a wide range of models containing
massless fermions and bosons.
In this paper, we concentrate on the chiral Schwinger model which
describes the quantum electrodynamics of massless electrons in two
dimensions (for this reason we will call it also ${\rm QED}_2$).
This choice is motivated by the physical relevance of the Schwinger
model in physics
\flat, \cmpl.
The usual approach to the Schwinger model is to integrate in the path
integral the fermionic degrees of freedom first. In this way, one obtains
an effective theory of massive gauge fields when nontrivial
topological sectors are absent \fp,
\ref\bnd{J. Barcelos-Neto, A. Das, {\it Phys. Rev.} {\bf
D33} (1986), 2262; {\it Zeit. Phys.} {\bf C32} (1986), 527.}.
However, in the case of nontrivial topological sectors, there is, at
least to our knowledge, no known way of integrating over the fermionic
degrees of freedom on a Riemann surface.
For this reason, we follow here another strategy,
eliminating first
the gauge
fields from the path integral. The idea behind this strategy is that the
physical fields in the Schwinger model are the fermions and not the
gauge fields. The latter have in fact only a discrete number of degrees of
freedom, provided by the zero modes and by the instantonic part
$A^{\rm cl}$.
After the integration over the quantum part of the gauge fields
$A^{\rm qu}$, we obtain indeed the effective field theory of
the massless electrons.
The electrons still remain minimally coupled to the instantonic and
harmonic components of the gauge fields, but these appear now only as
external fields. All the properties of the theory
which do not depend on the nontrivial
topological sectors, are instead concentrated in a nonlocal term,
describing the self-interactions between the fermions. One of these
properties, which is physically very important, is
the high energy behavior of the electrons.
This behavior is clearly not influenced by the
presence of external instantonic fields because it is entirely
dominated by the
the electromagnetic forces at very short distances.
As an upshot, we compute here explicitly the asymptotic form
of the potential governing the interactions between the electrons at
very high energies.
As we show, it strongly depends on the topology.
On a sphere, for example, the potential vanishes at very high energies, which
is a rather surprising result.
On a Riemann surface with genus $g\ge 1$, instead, the potential does
appearently not vanish at short distances due to the presence of zero modes.
Unfortunately, we are not able to estimate this zero mode
contribution exactly, but
from the physical point of view, it would be surprising that the short
distance behavior of the particles is determined by the zero modes.
Thus, we believe that also on a Riemann surface of genus $g\ge 1$
the electrons become free at high energies as it happens on the sphere.
Finally, the behavior of the potential on a Riemann surface is completely
different from that of the flat case, where the manifold has for
instance
the
topology of a disk.
In this sense, it would be very interesting to study the Schwinger
model also on the Euclidean space ${\rm\bf R}^2$. However, the procedure of
integrating in the gauge degrees of freedom followed here is no longer
possible on ${\rm\bf R}^2$ since there are problems in constructing the
propagator of the gauge fields as we pointed out in ref. \ferqed.
To conclude, we mention that the effective theory of the electrons found
here resembles very much the nonlocal generalization of the
Thirring model \ref\thirring{W. Thirring, {\it Ann. Phys. (N. Y.)} {\bf
3} (1958), 91.}
in the sense of ref. \ref\moffat{J. W.
Moffat, {\it Phys. Rev.} {\bf D41} (1990), 1177.} and we believe
therefore that it describes an integrable model.\smallskip
The material presented in this paper is divided as follows.
In Section 2 we show how the introduction of the Feynman gauge
simplifies the quantization of the abelian gauge field theories on a
manifold with conformally flat metric. In Section 3 we restrict
ourselves to the case of a general closed and orientable Riemann
surface.
An instantonic solution $A^{\rm cl}$
of the field equations in the Feynman gauge is
explicitly derived.
The first Chern class of the connection $A^{\rm cl}$ is $c_1=2\pi k$.
We show that $A^{\rm cl}$ is necessarily multivalued, but this
multivaluedness can be reabsorbed performing a gauge transformation
provided $k$ is an integer. In Section 4 we quantize the abelian gauge
fields $A^{\rm qu}$ representing small perturbations around the
instantonic solution $A^{\rm cl}$. Following ref. \ferqed, we derive the
explicit expression of the propagator of the $A^{\rm qu}$.
The relevant properties of the propagator, like orthogonality with
respect to the space of the flat connections and singlevaluedness around
the nontrivial homology cycles, are proven.
In this way, we can exactly
solve the pure abelian gauge field theory coupled to
an external current computing the generating functional of the
correlation functions on any nontrivial line bundle.
In Section 5 we treat the quantum electrodynamics (${\rm QED}_2$) on a
Riemann surface. The knowledge of the propagator and the fact that it is
orthogonal to the zero modes allows the integration
over the gauge fields $A^{\rm qu}$ in the path integral.
As a consequence, we obtain the effective action of electrons mentioned
above.
Finally, in Section 6 we evaluate the relativistic potential governing
the forces between the electrons at very short
distances.
Finally, we discuss in the Conclusions the possible extensions of the
results presented here.
\vfill\eject
\newsec{THE GAUGE FIXING CONDITION ON A RIEMANN SURFACE}
\vskip 1cm
In this Section we consider the Maxwell Field Theory (MFT) on a
Riemann surface $M$ coupled to an external source $J^\alpha$. This is a
pure gauge field theory with $U(1)$ gauge group of symmetry and the following
action:
\eqn\action{S_{\rm free}=\int_Md^2z\sqrt{g}\left({1\over
4}F_{\alpha\beta}F^{\alpha\beta} +J_\alpha A^\alpha\right)}
where $g={\rm det}|g_{\alpha\beta}|$.
$M$ is now a general, closed and orientable Riemann surface of genus
$g$, provided with an Euclidean metric $g_{\alpha\beta}$.
A covering of $M$ is given by a system of open sets $\{U_i\}$
parametrized by the local complex coordinates $z^{(i)}$ and $\bar
z^{(i)}$.
In the following we will drop the indices $i$ corresponding to the local
patches
of the covering.
Greek indices will denote complex indices. For instance $A_\alpha(z,\bar
z)\equiv(A_z(z,\bar z),A_{\bar z}(z,\bar z))$.
Moreover, let $\omega_i(z)dz$, $i=1,\ldots,g$, be a set of holomorphic
differentials, normalized in the following way with respect to the
canonical basis of homology cycles $A_i$ and $B_i$:
\eqn\homcycles{\oint_{A_i}\omega_jdz=0\qquad\qquad\qquad
\oint_{B_i}\omega_jdz=\Omega_{ij}}
$\Omega_{ij}$ is called the period matrix.
The tensor $F_{\alpha\beta}$ represents the usual field strength:
\eqn\fdstr{F_{\alpha\beta}=\partial_\alpha A_\beta-\partial_\beta
A_\alpha\qquad\qquad\qquad
F^{\alpha\beta}=g^{\alpha\gamma}g^{\beta\delta}F_{\gamma\delta}}
It is easy to check that the only nonvanishing components of the tensor
$F_{\alpha\beta}$ are $F_{z\bar z}=-F_{\bar z z}$.
The action \action\ is what we need to compute the
propagator of the gauge fields.
\smallskip
As a first step in order to quantize the MFT we have to choose a gauge
fixing. A convenient choice is the set of covariant gauges
defined by the condition:
\eqn\gaugefix{g^{\alpha\beta} \partial_\alpha A_\beta=0}
We notice here that for a general choice of the metric the usual
splitting of the fields in transverse and longitudinal components does
not make sense. The gauge fixed partition function in the Euclidean
space looks as follows:
\eqn\partfunct{Z_0[J]=\int DA_zDA_{\bar z}
{\rm exp}\left[-\left(S_{\rm
free}+S_{\rm gf}\right)\right]}
where:
\eqn\sgf{S_{\rm gf}={1\over
2\lambda}\int_Md^2zg^{\alpha\beta}g^{\gamma\delta}\partial_\alpha
A_\gamma\partial_\beta A_\delta}
We ignore for the moment the Faddeev$-$Popov term containing the
decoupled ghost action.\smallskip
Due to the presence of the metric, the computation of the partition
function $Z_0[J]$ and therefore of the propagator becomes involved.
Still we can simplify $S_{\rm free}$ using the following observation:
\eqn\obs{{1\over
4}g^{\alpha\beta}g^{\gamma\delta}F_{\alpha\gamma}F_{\beta\delta}=
{1\over 2} g^{-1}F_{z\bar z}^2}
where $g^{-1}=g^{zz}g^{\bar z\bar z}-g^{z\bar z}g^{\bar zz}$.
In this way:
\eqn\sfree{S_{\rm free}={1\over 2}\int_Md^2zg^{-{1\over 2}}F_{z\bar z}^2}
However, the gauge fixing
part of the action $S_{\rm gf}$ remains complicated.
For this reason, we exploit the fact that on a two
dimensional manifold it is always possible to choose a conformally flat
metric of the kind:
\eqn\flat{g_{zz}=g_{\bar z\bar z}=0\qquad\qquad\qquad g_{z\bar
z}=g_{\bar zz}\qquad\qquad\qquad g^{z\bar z}=1}
In this metric $\sqrt{g}=g_{z\bar z}$ and
the gauge fixing condition of eq. \gaugefix\ becomes independent on
$g_{\alpha\beta}$:
\eqn\gf{\partial_z A_{\bar z}+\partial_{\bar z}A_z=0}
Moreover, the transversal
and longitudinal components of the fields can be defined through the
Hodge decomposition, as we will see in the next Section.\smallskip
Also with the above simplifications an easy solution of
the partial differential
equations that determine the components of the propagator is not
possible
unless we choose the following particular values of the parameter $\lambda$ in
eq. \sgf:\medskip
\item{a)} $\lambda=1$ (Feynman gauge).
\medskip
\item{b)} $\lambda=0$ (Lorentz gauge).\medskip
The advantages and drawbacks of the choice of the Landau gauge have
already been discussed in the Introduction and in \ferqed.
In order to formulate the interacting MFT in the simplest possible way
that easily generalizes to the Yang$-$Mills field theories, we
investigate now the theory in the Feynman gauge.\smallskip
\vskip 1cm
\newsec{THE CLASSICAL EQUATIONS OF MOTION}
\vskip 1cm
After an easy calculation, we find the following
expression for the total action $S=S_{\rm free}+S_{\rm gf}$
appearing in eq. \partfunct\ in the Feynman gauge:
\eqn\totalaction{S_{\rm MFT}
(A,J)=\int_Md^2z\left[g^{z\bar z}\left(\partial_z A_{\bar
z}\partial_zA_{\bar z}+
\partial_{\bar z} A_z\partial_{\bar z}A_z\right)+J_zA_{\bar z}+J_{\bar
z}A_z\right] }
The classical equations of motion related to this action
become now relatively simple:
\eqn\classeq{\partial_{\bar z}g^{z\bar z}\partial_{\bar z}A_z=J_{\bar z}
\qquad\qquad\qquad
\partial_zg^{z\bar z}\partial_zA_{\bar z}=J_z}
In order to study eq. \classeq, we split the fields using
the Hodge decomposition:
\eqn\hodgedec{A_\alpha=\epsilon_{\alpha\beta}\partial^\beta\varphi+
\partial_\alpha\rho+A_\alpha^{\rm har}+A_\alpha^I}
where $\epsilon_{z\bar z}=-\epsilon_{\bar z z}=ig_{z\bar z}$ is the
completely antisymmetric Levi$-$Civita tensor and $i^2=-1$.
\smallskip
In eq. \hodgedec\ $\varphi(z,\bar z)$ is a real scalar field
representing the transverse degree of freedom. As a matter of fact, the
components of the gauge fields in $\varphi$ are $A^{\rm
T}_z=\partial_z\varphi$ and
$A^{\rm
T}_{\bar z}=-\partial_{\bar z}\varphi$ and it is easy to check that they
satisfy eq. \gf, so that they are purely transversal.
The component of the gauge fields in the real scalar field
$\rho(z,\bar z)$
denote instead the longitudinal degrees of freedom.
Finally, the
$A_\alpha^{\rm har}$ represent the flat connections in the abelian case.
They take into account of the $g$ zero modes
$\omega_i(z)dz$ discussed in the previous Section.
We notice that the zero modes describe nonphysical degrees of freedom
and therefore, in the interacting case, they should not propagate within
the amplitudes.\smallskip
The first three terms of the right hand side of eq. \hodgedec\
correspond to the exact, coexact and harmonic forms in which a
differential $A_\alpha$ can be decomposed on a Riemann surface. This is
the singlevalued part of the gauge fields, while $A_\alpha^{\rm I}$ is
an instantonic and periodically multivalued differential.
The multivaluednes occurs when the field $A_\alpha^{\rm I}$ is
transported along a nontrivial homology cycle of the Riemann
surface.
Remembering that the first Chern class is defined by
\eqn\fchern{c_1=\int_Md^2zF_{z\bar z}}
with $c_1=2\pi k$ and $k\in {\rm\bf Z}$, we can say that
$A_\alpha^{\rm I}$ is a solution of these equations defined on a
nontrivial line bundle $P(M,U(1))$ characterized by $c_1\ne 0$.
\smallskip
The point of view we will take here in quantizing the Maxwell field
theory is that of small perturbations around an instantonic classical
field $A_\alpha^{\rm cl}\equiv A_\alpha^{\rm I}$.
Let us now compute the explicit form of $A_\alpha^{\rm I}$.
These $A_\alpha^{\rm I}$ are very similar to the Maxwell connections of
ref. \thom, apart from the fact that they do not
satisfy the pure Maxwell equations, but the
gauge fixed Maxwell equations \classeq.
In order to compute $A_\alpha^{\rm I}$, we have to consider
the homogeneous classical equations of
motion putting $J_\alpha=0$ in eq. \classeq.
Due to the fact that the Riemann
surface $M$ is a compact, orientable manifold without boundary,
the only possible nontrivial solutions of the
homogeneous equations of motion are:
\eqn\internal{
\partial_{\bar z}A_z^{\rm I}=\alpha_1g_{z\bar z}\qquad\qquad\qquad
\partial_zA_{\bar z}^{\rm I}=\alpha_2g_{z\bar z}}
$\alpha_1$ and $\alpha_2$ being arbitrary constants.
However, $\alpha_1$ and $\alpha_2$ become uniquely determined once we
require that $A_\alpha^{\rm I}$ satisfies eq. \fchern\ and the gauge
condition \gf.
As a matter of fact, substituting eq. \internal\ in eq. \fchern, we have:
\eqn\firstcond{\alpha_1-\alpha_2={2\pi k\over A}}
where $A=\int_Md^2zg_{z\bar z}$ is the total area of the Riemann surface
$M$. Finally, exploiting eq. \gf\ we get:
\eqn\seccond{\alpha_1+\alpha_2=0}
{}From eqs. \firstcond\ and \seccond\ we get the following implicit
expression for the instantonic solutions:
\eqn\instant{\partial_{\bar z}A_z^{\rm I}
={\pi k\over A}g_{z\bar z}\qquad\qquad\qquad
\partial_zA_{\bar z}^{\rm I}=-{\pi k\over A}g_{z\bar z}}
Let us notice that in this way we have obtained nontrivial instantonic
solutions of the Maxwell field theory in the Feynman gauge.
Their first Chern class is $2\pi k$.\smallskip
In order to have the explicit expression of $A_\alpha^{\rm I}$,
we try the ansatz:
\eqn\soljeden{A_z^{\rm I}=\alpha_1\int_M d^2w\partial_z\tilde K(z,w)g_{w\bar
w}-\sum\limits_{i,j=1}^g\alpha_1A \omega_i(z)\left| Im\enskip
\Omega_{ij}\right|^{-1} \int_{\bar z_0}^{\bar z}\bar\omega_j(\bar
w)d\bar w}
\eqn\soldwa{A_{\bar z}^{\rm I}=\alpha_2\int_M d^2w\partial_{\bar z}
\tilde K(z,w)g_{w\bar
w}-\sum\limits_{i,j=1}^g\alpha_2A \bar\omega_i(\bar z)\left|Im\enskip
\Omega_{ij}\right|^{-1} \int_{z_0}^z\omega_j(w)dw}
where
\eqn\ktilde{\tilde K(z,w)={\rm log}|E(z,w)|^2+\sum\limits_{i,j=1}^g
\left[{\rm
Im} \int_{w}^z\omega_j(s)ds\right]
\left|Im\enskip
\Omega_{ij}\right|^{-1}
\left[{\rm
Im} \int_{w}^z\omega_j(s)ds\right]}
Remembering that
\eqn\ktildedef{\partial_z\partial_{\bar z}\tilde K(z,w)=
\delta^{(2)}_{z\bar z}(z,w)+\omega_i(z)\left|Im\enskip
\Omega_{ij}\right|^{-1}\bar\omega_j(\bar z)}
we can easily check that the solutions \soljeden\ and \soldwa\ satisfy
exactly the eqs. \instant.
The fields $A_\alpha^{\rm I}$ defined in \soljeden\ and \soldwa\ are
periodic around the homology cycles $A_i$ and $B_i$, $i=1,\ldots,g$.
Transporting $n_l$ times $A_\alpha^{\rm I}$ around $B_l$,
$l=1,\ldots,g$,
we get for example:
\eqn\gtone{A_z^{\prime\rm I}=A_z^{\rm I}+2\pi
kn_l\sum\limits_{i,j=1}^g
\bar\Omega_{lj}(\Omega-\bar \Omega)^{-1}_{ji}\omega_i(z)}
\eqn\gtbone{A_{\bar z}^{\prime\rm I}=A_{\bar z}^{\rm I}-2\pi
kn_l\sum\limits_{i,j=1}^g
\Omega_{lj}(\Omega-\bar \Omega)^{-1}_{ji}\bar\omega_i(\bar z)}
This periodicity amounts however to a gauge transformation.
In fact, let us set
\eqn\uone{
{g(z,\bar z)\over g(z_0,\bar z_0)}={\rm exp}
\left[
2\pi ikn_l\sum\limits_{i,j=1}^g
\left(
\int_{z_0}^z
\bar\Omega_{lj}
(\Omega-\bar \Omega)^{-1}_{ji}
\omega_i(w)dw-
\int_{\bar
z_0}^{\bar z}
\Omega_{lj}
(\Omega-\bar \Omega)^{-1}_{ji}
\bar \omega_i(\bar w)d\bar w
\right)
\right]
}
We claim that $g(z,\bar z)$ represents a singlevalued
$U(1)$  gauge transformation on a Riemann surface corresponding to the
flat line bundle given by the holomorphic connections
$A_{\alpha,i}^{\rm har}=A^{\prime {\rm I}}_\alpha-A^{\rm I}_\alpha$.
As a matter of fact, the connection contained in the exponent of eq. \uone\
is of the form
\eqn\flatconn{2\pi k n_l\sum\limits_{i,j=1}^g
\bar\Omega_{lj}(\Omega-\bar \Omega)^{-1}_{ji}\omega_i(z)+{\rm
c.c.}=\alpha_kn_l}
$\alpha_i$ being the real harmonic differential with the following
holonomies around the homology cycles:
$$\oint_{A_i}\alpha_j=\delta_{ij}\qquad\qquad\qquad
\oint_{B_i}\alpha_j=0$$
If we consider the behavior of \soljeden\ and \soldwa\
around the homology cycles $A_i$, we arrive to
an analogous result in which the
$\alpha_i$ are replaced by the real harmonic differentials $\beta_i$.
Using eq. \flatconn\ in eq. \uone, it is now easy to see that
$g(z,\bar z)$  has exactly the form of a good $U(1)$ gauge
transformation
according to ref. \ref\amv{L. Alvarez-Gaum\'e,
G. Moore, C. Vafa, {\it Comm. Math.
Phys.} {\bf 106} (1986), 1.}.
The proof that eq. \gtone\ and \gtbone\ represent gauge transformations
is concluded noting that these equations
can be rewritten in the following form:
\eqn\ggproof{A_\alpha^{\prime\rm I}=A_\alpha^{\rm
I}+ig^{-1}\partial_\alpha g}
Of course, the above gauge transformation is well defined only if $k$ in eq.
\fchern\ is an integer, otherwise $g(z,\bar z)$ becomes multivalued.
Therefore, we have proven that the gauge connection $A_\alpha^{\rm I}$
given in \soljeden\ and \soldwa\ are the correct solutions of the
equations \instant. They represent the generalization to a Riemann surface
of the connections given in the case of a torus
in refs. \thom\ and \wipf. In both cases the connections
are periodic around the homology cycles but the
periodicity amounts to a gauge transformation.
\vskip 1cm
\newsec{THE PROPAGATOR IN THE FEYNMAN GAUGE}
\vskip 1cm
In this Section we construct the propagator
\eqn\propdef{G_{\alpha\beta}(z,w)\equiv\langle A_\alpha^{\rm qu}
(z,\bar z)A_\beta^{\rm qu}(w,\bar
w)\rangle}
{}From the action \totalaction\ it is easy to see that the components
$G_{z\bar w}(z,w)$ and $G_{\bar z w}(z,w)$ vanish identically. The
remaining components of the propagator satisfy the following two
equations:
\eqn\propone{\partial_{\bar z}g^{z\bar z}\partial_{\bar
z}G_{zw}(z,w)=\delta_{\bar zw}(z,w)+{\rm zero}\enskip{\rm modes}}
\eqn\proptwo{\partial_zg^{z\bar z}\partial_zG_{\bar z\bar w}(z,w)=
\delta_{z\bar w}(z,w)+{\rm zero}\enskip{\rm modes}}
Analogous equations are to be solved in the variable $w$.
The exact form of the zero mode contribution appearing in the left hand
side of eqs. \propone\ and
\proptwo\ will be given below.\smallskip
In solving eqs. \propone\ and \proptwo\ we can use the techniques
developed in ref. \ferqed\ in the case of the Lorentz gauge fixing.
The only difference that occurs in the Feynman gauge is that now both
the transversal and longitudinal fields are propagated, so that we have
two distinct equations of motion in $A_z$ and $A_{\bar z}$.
In the Lorentz gauge, instead, there is only one equation in $A_z$ or,
equivalently, in $A_{\bar z}$ (see ref \ferqed\ for more details).
A detailed account of the way in which the components of the propagator
can be found was already provided in ref. \ferqed. Here we just give the
result:
\eqn\gzw{G_{zw}(z,w)=-\int_Md^2tg_{t\bar t}\partial_zK(z,t)\partial_wK(w,t)}
\eqn\gzbwb{G_{\bar z\bar w}(z,w)=-\int_Md^2tg_{t\bar t}
\partial_{\bar z}K(z,t)\partial_{\bar w}K(w,t)}
where $K(z,w)$ is the usual scalar Green function defined by the relations:
\eqn\scalone{
\partial_z\partial_{\bar z}K(z,w)=\delta_{z\bar z}^{(2)}(z,w)-
{g_{z\bar z}\over A}}
\eqn\scaltwo{
\partial_{\bar z}\partial_wK(w,z)=-\delta^{(2)}(z,w)+\bar\omega_i(\bar
z)\left[{\rm Im}\enskip \Omega\right]^{-1}_{ij}\omega_j(w)}
\eqn\scalthree{\int_Md^2t g_{t\bar t}K(z,t)=0}
As it is well known, $K(z,w)$ has an explicit expression in terms of the
Green function \ktilde:
\eqn\kzw{K(z,w)=\tilde K(z,w)-{1\over A}\int d^2t g_{t\bar t}\left(\tilde
K(z,t)+\tilde K(t,w)\right)+{1\over A^2}\int\int d^2sd^2t\enskip
g_{t\bar
t}g_{s\bar s}\tilde K(t,s)}
Now we discuss the properties of the Green functions \gzw\ and \gzbwb.
In particular, we show that they fulfill all the possible requirements
of a physical propagator.
First of all, let us check that the components $G_{zw}(z,w)$ and
$G_{\bar z\bar w}(z,w)$
of the propagator
given in eq. \gzw\ and \gzbwb\ satisfy the correct equations of motion
\propone\ and \proptwo.
Using the properties \scalone\ and \scalthree\ of the scalar Green
function $K(z,w)$ it is possible to show that:
\eqn\fcheck{\partial_{\bar z}g^{z\bar
z}\partial_{\bar z}G_{zw}(z,w)=\delta^{(2)}(z,w)-
\sum\limits_{i,j=1}^g\bar\omega_i(\bar
z)\left[{\rm Im}\enskip \Omega\right]^{-1}_{ij}\omega_j(w)}
An analogous equation can be derived for the component $G_{\bar z\bar
w}(z,w)$ of the propagator.
In this way we have determined the contribution of the zero modes in the
right hand sides of eqs. \propone\ and \proptwo.
Moreover, the propagator is singlevalued along the nontrivial homology cycles
of the Riemann surface:
\eqn\singval{\oint_\gamma dz G_{zw}(z,w)=\oint_\gamma dw G_{zw}(z,w)=
\oint_\gamma d\bar z G_{\bar z\bar w}(z,w)=\oint_\gamma d\bar w G_{\bar
z\bar w}(z,w)=0}
where $\gamma=A_i,B_i$, $i=1,\ldots,g$.
This is a consequence of the singlevaluedness of the scalar Green
function \kzw.
Eq. \singval\ guarantees that the fields
$A_\alpha^{\rm qu}$ are singlevalued quantum perturbations on $M$.
Finally, $G_{\alpha\beta}(z,w)$ is orthogonal to the space of harmonic
connections $A_\alpha^{\rm har}$.
This is a trivial consequence of the fact that
$$\int d^2z\partial_zK(z,t)\bar\omega_i(\bar z)=
\int d^2z\partial_{\bar z}K(z,t)\omega_i(z)=0$$
for $i=1,\ldots,g$.
Due to the above equation, it is easy to prove that
for any harmonic differential $A^{\rm har}_\alpha$
we have:
\eqn\orthogprop{\int d^2z A^{\rm har}_{\bar z}\int d^2w G_{zw}(z,w)J_{\bar w}=
\int d^2z A^{\rm har}_z\int d^2w G_{\bar z\bar w}(z,w)J_w=0}
for $i=1,\ldots,g$.
Eq. \orthogprop\ implies that $G_{\alpha\beta}(z,w)$ does not propagate
the zero modes inside of the amplitudes. This can be explicitly seen
considering the
solutions of the classical equations of motion \classeq:
\eqn\potentials{A_w(w,\bar w)=\int d^2 z G_{zw}(z,w)J_{\bar
z}\qquad\qquad\qquad
A_{\bar w}(w,\bar w)=\int d^2 z G_{\bar z\bar w}(z,w)J_z}
At this point, we split the current $J_\alpha$ as follows:
\eqn\jazzb{J_z(z,\bar z)=\partial_z\chi(z,\bar z)+ia_i
[Im\enskip
\Omega]^{-1}_{ij}\omega_j(z)}
where
\eqn\auxchang{a_i=-i\int d^2z\bar\omega_i(\bar z)J_z(z,\bar z)}
and $\chi(z,\bar z)$ is a complex scalar field.
The component $J_{\bar z}$  can be found taking the complex conjugate in
the right hand side of eq. \jazzb.
This decomposition is equivalent to the Hodge decomposition \hodgedec\
apart from the absence of the instantonic fields.
Now we insert the gauge field $A_{\bar w}$ given in eqs. \potentials\
in the original equations of motion \classeq.
Using eq. \fcheck\ we get:
\eqn\resultone{\partial_w g^{w\bar w}\partial_w\int d^2z G_{\bar z \bar
w}(z,w) J_z(z,\bar z)=\partial_w \chi(w,\bar w)}
\eqn\resultwo{\partial_{\bar w} g^{w\bar w}\partial_{\bar w}\int
d^2z G_{zw}(z,w) J_{\bar z}(z,\bar z)=\partial_{\bar w} \bar\chi(w,\bar w)}
showing that the zero modes are not propagated by the
propagator.\smallskip
A last property of $G_{\alpha\beta}(z,w)$ comes from the fact that the
Hodge decomposition \hodgedec\ is not invertible if $\varphi$ and $\rho$
are constants. This implies the following condition on $\varphi$ and
$\rho$:
$$\int_Md^2z\sqrt{g}\varphi(z,\bar z)=\int_Md^2z\sqrt{g}\rho(z,\bar
z)=0$$
In agreement, the biharmonic Green function
\eqn\biharm{G(z,w)=\int_Md^2zg_{t\bar t}K(z,t)K(w,t)}
from which the propagator $G_{\alpha\beta}(z,w)$ can be evaluated,
should satisfy the relations:
$$\int_Md^2z\sqrt{g}G(z,w)=\int_Md^2w\sqrt{g}G(z,w)=0$$
Exploiting eq. \scalthree\
it is easy to see that the above equations hold.
\vskip 1cm
\newsec{TWO DIMENSIONAL QUANTUM ELECTRODYNAMICS ON A RIEMANN SURFACE}
\vskip 1cm
Until now, we have investigated the classical MFT in the presence of an
external current.
In this section we consider the
case in which the gauge fields are allowed to interact with other matter
fields.
In particular, we treat the chiral Schwinger model \sch\ discussed also in
\ferqed\ in the case of trivial line bundles.
Tha action of the model is given by:
$$S_{{\rm QED}_2}[A,\bar\psi,\psi,\bar\xi,\xi]=
S_{\rm MFT}(A,J=0)+\int_Md^2z\left[
\bar \psi_\theta(\partial_{\bar z}+A_{\bar z})
\psi_\theta+\bar \psi_{\bar \theta}(\partial_z+A_z)\psi_{\bar \theta}+\right.
$$
\eqn\actqed{\left.J_zA_{\bar z}+J_{\bar z}A_z+g_{\theta\bar \theta}(\xi_\theta
\psi_{\bar \theta}+\xi_{\bar \theta}\psi_\theta)+
g_{\theta\bar\theta}(\bar\xi_\theta\bar\psi_{\bar\theta}+\bar\xi_{\bar\theta}
\bar\psi_\theta)\right]}
where $\xi$, $\bar\xi$ are the external currents related to the fields $\psi$,
$\bar\psi$ respectively and $g_{\theta\bar\theta}=\sqrt{g_{z\bar z}}$.
For simplicity, we assume that the coupling constant between gauge
fields and spinors is equal to one.
Moreover, we do not need the external currents associated to the gauge
fields, so that they are set to zero.
The spinor indices in eq. \actqed\ are denoted with $\theta$, $\bar\theta$.
We assume that the ``physical" boundary conditions for $\psi$ and $\bar\psi$
when transported
along the homology cycles on $M$ are given by
the even spin structure $s=\left[\matrix{\vec a_0 \cr \vec b_0\cr}\right]$.
$\vec a_0$ and $\vec b_0$ are two vectors of dimension g whose elements are
half integers such that $4\vec a_0\cdot \vec b_0=0$ mod $2$
\ref\fay{J. D. Fay, {\it Lect. Notes in Math.} {\bf 352}, Springer
Verlag, 1973.}.
The matter currents will be denoted as follows:
\eqn\currents{J_z^m=\bar\psi_\theta \psi_\theta\qquad\qquad\qquad J_{\bar
z}^m=\bar\psi_{\bar \theta} \psi_{\bar\theta}}
These currents do not contain zero modes with respect to the operators
$\partial_\alpha$, since the even spin structures
do not admit holomorphic sections.
Since we are considering small perturbations around the instantonic
solutions of the equations of motion $A_\alpha^{\rm I}$, we
split in eq. \actqed\ the gauge fields into a
quantum part and a classical and topologically nontrivial
contribution given by eqs. \instant:
\eqn\splitting{A_z=A_\alpha^{\rm qu}+A^{\rm I}_\alpha}
where $A_\alpha^{\rm
qu}=\epsilon_{\alpha\beta}\partial^\beta\varphi+\partial_\alpha
\rho+A_\alpha^{\rm har}$.
The advantage of the splitting \splitting\ is that
$A_\alpha^{\rm I}$
satisfies the classical equations of motion and eq. \instant.
Hence, apart from a constant that can be factored out in the path
integral, it is easy to see that the only dependence on $A_\alpha^{\rm I}$
of \totalaction\
is in the term with the external currents which
has explicitly been neglected in eq. \actqed. Moreover, also the
harmonic components of the gauge fields do not contribute to $S_{\rm
MFT}(A^{\rm qu},J=0)$.
As a consequence, the generating functional $Z[\bar \xi,\xi]$ of ${\rm
QED}_2$ becomes:
$$Z_{\rm QED}[\bar \xi,\xi]=\int D\tilde A_\alpha^{\rm qu}D\bar\psi
D\psi \prod\limits_{i=1}^gd\theta_id\phi_i{\rm exp}\left\{-\left[S_{\rm
MFT}(\tilde A^{\rm qu},J=0)+\int_Md^2zA_\alpha^{\rm qu}J^{m,\alpha}
\right.\right.$$
\eqn\partqed{+\left.\left.\int_Md^2z\left(A_\alpha^{\rm har}J^{m,\alpha}+
\bar \psi_\theta(\partial_{\bar z}+A_{\bar z}^{\rm I})
\psi_\theta+\bar \psi_{\bar \theta}(\partial_z+A_z^{\rm I})\psi_{\bar \theta}
+\bar\psi^{\bar\theta}\xi_{\bar\theta}+\bar\psi^\theta\xi_\theta+
\psi^{\bar\theta}\bar\xi_{\bar\theta}+\psi^\theta\bar\xi_\theta
\right)\right]\right\}}
where $\tilde A^{\rm qu}=A^{\rm qu}-A^{\rm har}$.
In the above equations we have parametrized the space of flat
connections in the usual way \amv:
$$A_z^{\rm har} dz+A_{\bar z}^{\rm har}d\bar z=
2\pi i({\bf \phi}+\bar\Omega{\bf \theta})\cdot(\Omega-\bar
\Omega)^{-1} \cdot \bar \omega(z)dz+{\rm c.c}$$
Therefore, the integration over the parameters $\theta_i$ and $\phi_i$ is
a sum over the flat connections.
We note that the only dependence on $A^{\rm har}_\alpha$ is in the
interaction term $\int_Md^2zA_\alpha^{\rm har}J^{m,\alpha}$.
Moreover, it is important to stress that the fields $A_\alpha^{\rm I}$
are singlevalued on $M$, apart from a gauge transformation which can be
reabsorbed by a gauge transformation of the spinor fields in such a way
that the whole action in eq. \partqed\ remains singlevalued.\smallskip
The functional $Z_{\rm QED}[\bar \xi,\xi]$ can be further simplified.
One way to do this, is to perform an integration over $\bar\psi$ and
$\psi$. However, this yields a free theory of massive vector
fields\foot{Let us notice that it is not clear if the Schwinger model
on a manifold is simply a covariantized version of the model in the flat
case, see e.g. \bnd\ and
\ref\leutwyler{ H. Leutwyler, {\it Phys. Lett} {\bf 153B}
(1985), 65.}.},
for which it is not easy to compute the amplitudes on a Riemannn surface.
Most important, these amplitudes have no direct physical significance.
In fact, the physical fields are represented by the fermions, while the
gauge fields are not observable and have only a discrete number of
degrees of freedom in two dimensions.
{}From this point of view, it seems preferable to
eliminate the gauge fields from eq. \partqed.
To this purpose, we have to evaluate the following path integral:
\eqn\partatqu{\tilde Z^{\rm qu}[J^m]=\int D\tilde A^{\rm qu}{\rm
exp}\left[ -\left(S_{\rm MFT}(\tilde A^{\rm qu},J=0)+\int_M\tilde
A_\alpha^{\rm
qu}J^{m,\alpha}\right)\right]}
We note that $\tilde Z^{\rm qu}[J^m]$ contains only
a sum over exact and coexact differentials, since the harmonic
components have already been extracted.
Hence, the operators
$\partial_\alpha$ in the exponent of
eq. \partatqu\ act on a the space of differentials
which is orthogonal to the harmonic components and we are free to
integrate by parts.
Using standard techniques we perform in eq. \partatqu\
the change of variables:
\eqn\chgvar{\tilde A_\alpha^{\prime {\rm qu}}=\tilde A_\alpha^{\rm qu}+
{1\over 2}\int d^2wg_{w\bar w}
G_{\beta\alpha}(z,w)J^{m,\beta}(w,\bar w)}
where $\alpha=z,\bar z$ and $\beta=w,\bar w$.
Again, $\tilde A^{\prime\rm qu}$
is still orthogonal to the space of the harmonic
differentials because of the properties of the propagator explained in
the previous section.
Substituting \chgvar\ in eq. \partatqu, we obtain
$$Z_{\rm QED}[\bar\xi,\xi]=\int D\bar\psi
D\psi\prod\limits_{i=1}^gd\theta_id\phi_i
{\rm
exp}\left\{-\left[\int_Md^2z\left[\bar\psi_\theta(\partial_{\bar
z}+A^I_{\bar z})\psi_\theta+
\bar\psi^\theta\xi_{\theta}+\psi^\theta\bar\xi_\theta
\right]\right.\right.$$
\eqn\zqedfin{\left.\left.
+\int_Md^2zg_{z\bar z}A_\alpha^{\rm har}J^{m,\alpha}+
{1\over4}\int_Md^2zd^2z'\left(J_z^m(z,\bar z)G_{\bar
z'\bar z}(z,z')J^m_{z'}(z',\bar z')\right)+{\rm
c.c}\right]\right\}}
Therefore, the generating functional of ${\rm QED}_2$ on a Riemann surface $M$
can be expressed as a theory of fermions with a self-interacting potential
which is given by the two point function of the gauge fields.
The latter is explicitly given in eqs. \gzw\ and \gzbwb, so that it is
possible,
at least in principle, to compute the behavior of this potential at very high
energies, i.e. in the limit $z\rightarrow z'$.\smallskip
Notice that the formula \zqedfin\ has been obtained only because the
propagator $G_{\alpha\beta}(z,z')$ on a Riemann surface does exist and
it is explicitly known. In this respect, difficulties may arise when $M$
is a noncompact manifold, the complex plane included. In this case, in
fact, the propagator of the gauge fields satisfying the physical
boundary conditions is not easy to construct \ferqed.
The problem is that the biharmonic Green function \biharm\ with the
desired boundary conditions from
which we can derive $G_{\alpha\beta}(z,w)$ as explained in \ferqed\ does
not exist on noncompact manifolds
\ref\snwc{L. Sario, M. Nakai, C. Wang, L. O. Chung,
{\it Lect. Notes in Math.} {\bf 605}, Springer Verlag, 1977.}.
\vskip 1cm
\newsec{HIGH ENERGY BEHAVIOR OF QUANTUM ELECTRODYNAMICS}
\vskip 1cm
In this section we compute the behavior of the potential
$G_{\alpha\beta}(z,w)$ appearing in eq. \zqedfin\ at short distances.
First of all, we study the flat case, namely a disk $B$ of unitary radius.
On the disk, the propagator of the gauge fields is given by the
derivatives of the following biharmonic Green function:
\eqn\biharmonic{G(z,w)={1\over 2}|z-w|^2{\rm log}\left[{|z-w|^2\over
(1-\bar w z)(1-w\bar z)}\right]+{1\over 2}(|z|^2-1)(|w|^2-1)}
satisfying the biharmonic equation
$$(\partial_{\bar z}\partial_w)^2G(z,w)=\delta^{(2)}(z,w)$$
on $B$.
For example, $G_{zw}(z,w)\equiv \partial_z\partial_wG(z,w)$ reads as
follows:
\eqn\gzwdisk{
G_{zw}(z,w)
={1\over 2}
\left[
-\left(
{\bar z-\bar w\over
z-w}
\right)
 +\bar z\left(
{\bar z-\bar w\over 1-w\bar z}
\right)
+\bar w\left(
{\bar z-\bar w\over 1-\bar w z}
\right)+\bar z\bar w\right]}
Analogously, one can compute $G_{\bar z\bar w}(z,w)\equiv \partial_{\bar
z}\partial_{\bar w}G(z,w)$.
The correct boundary conditions of the fields at the boundary $\partial
B$ of $B$ are:
\eqn\boundcond{A_{\bar z}=A_z=0\qquad\qquad\qquad z,\bar z\in \partial
B}
In this way, the spurious harmonic gauge transformation typical of the
covariant gauge fixing \gf\ are eliminated.
It is easy to show that the propagator \gzwdisk\ vanishes at the boundary
in $z$ and $w$ separately according to eq. \boundcond.
Moreover, if \boundcond\ is satisfied, the Hodge decomposition
\hodgedec\ is still valid (see e.g. \fp).
As previously remarked, there are no harmonic components in this case.
We remember also that in order to derive eq. \zqedfin\ we used the
freedom of doing partial integrations in the action \actqed.
All the possible boundary terms that can be generated on $B$ in this way
are however killed by the boundary conditions \boundcond, so that eq.
\zqedfin\ holds also on a disk.
Now we compute the short distance behavior of $G_{zw}(z,w)$. Setting
$z-w=\rho e^{i\theta}$ with $\rho\rightarrow 0$, we get from eq.
\gzwdisk:
\eqn\sdbdisk{\lim_{{z\to w}\atop {\bar z\to\bar
w}}G_{zw}(z,w)=-{1\over 2}e^{-2i\theta}+{1\over 2}\bar w^2}
A complete different result arises in the case of a sphere $S^2$ with
metric $g_{z\bar z}dzd\bar z={dzd\bar z\over (1+z\bar z)^2}$.
On the sphere, the function $K(z,w)$ of eq. \kzw\ becomes:
\eqn\kzwstwo{K(z,w)={\rm log}\left[{|z-w|^2\over (1+z\bar z)(1+w\bar
w)}\right]}
In order to find the short distance behavior of $G_{zw}(z,w)$ on $S^2$
we have to compute:
\eqn\gzzstwo{G_{zw}(z,z)=
\int_{S^2}d^2t\left(\partial_zG(z,t)\right)^2g_{t\bar
 t}}
It is now possible to prove the following identity:
$$\left(\partial_zK(z,t)\right)^2=-(\partial_z+g_{z\bar z}\partial_z
g^{z\bar z})\partial_zK(z,t)$$
Exploiting the above equation and the fact that
$\int_{S^2}d^2t\partial_zG(z,t)g_{t\bar t}=0$, we find:
\eqn\sdbstwo{G_{zw}(z,z)=0}
This result is profoundly different with respect to that of eq.
\sdbdisk\ and shows how the topology can influence the behavior of the
fermions at high energy.
In particular, eq. \sdbstwo\ shows that at short distances the electrons
do not feel any interaction on the sphere.\smallskip
What happens in the case of a Riemann surface?
We expect a result similar to that of the sphere, with the only
difference that now the right hand side of eq. \sdbstwo\ will be not
zero due to the presence of the zero modes.
The evaluation of
\eqn\gzzm{G_{zz}(z,z)=\int_Md^2t(\partial_zK(z,t))^2g_{t\bar t}}
requires an equation that expresses the function $(\partial_zK(z,t))^2$
in terms of linear combinations of $K(z,w)$ and its derivatives as we
did in the case of the sphere.
An equation of this kind has been derived in ref. \ref\ks{S. M. Kuzenko,
O. A. Solov'ev, {\it JETP Lett.} {\bf 51} (1990), 233.}.
Here we will only show that the integral \gzzm\ is proportional to a
zero mode following ref. \ref\ds{M. J. Dugan, H. Sonoda, {\it Nucl.
Phys.} {B289} (1987), 227.}.
First of all, instead of $\partial_zK(z,t)$ we consider the tensor:
\eqn\mzt{m_z(z,t)=\partial_zK(z,t)+\sum\limits_{i,j=1}^g\omega_i(z)\left|{\rm
Im}\enskip\Omega_{ij}\right|^{-1}\int_{\bar z_0}^{\bar z}\bar
\omega_j(\bar s)d\bar s}
This tensor is multivalued on the Riemann surface M but has the
advantage that it is easier to handle than $\partial_zK(z,t)$.
It is possible to see that in terms of $m_z(z,t)$ eq. \gzzm\ becomes:
\eqn\newexkzt{\int d^2tg_{t\bar t}(\partial_zK(z,t))^2=\int d^2t
g_{t\bar t}m_z^2(z,t)-A\left(
\sum\limits_{i,j=1}^g\omega_i(z)\left|{\rm
Im}\enskip\Omega_{ij}\right|^{-1}\int_{\bar z_0}^{\bar z}\bar
\omega_j(\bar s)d\bar s\right)^2}
Here we have exploited the fact that from eq. \scalthree\ it descends
that:
\eqn\scalfour{\int d^2tg_{t\bar t}\partial_zK(z,t)=0}
Now we evaluate $\int d^2tg_{t\bar t}m_z^2(z,t)$.
To do this, we expand $m_z(z,t)$ around the singularity at $z=t$:
\eqn\expansion{m_z(z,t)\sim{1\over z-t}-\Omega_z(z)+\omega_z(z)+O(z-t)}
where
\eqn\Omegaz{\Omega_z(z)={1\over A}\int_Md^2sg_{s\bar s}\partial_z\tilde
K(z,s)} and $\omega_z(z)$ is an irrelevant zero mode.
The expansion \expansion\ is motivated by the fact that $m_z(z,t)$
satisfies the following equation:
\eqn\eqdzb{\partial_{\bar z}m_z(z,t)=\delta^{(2)}_{z\bar
z}(z,t)-{g_{z\bar z}\over A}+\sum\limits_{i,j=1}^g\omega_i(z)\left|{\rm
Im}\enskip\Omega_{ij}\right|^{-1}\bar
\omega_j(\bar z)}
In fact, applying the operator $\partial_{\bar z}$ to the right hand
side of equations \expansion, we obtain a perfect agreement with eq.
\eqdzb.
At this point, it is possible to estimate also the expansion of $m^2_z(z,t)$
at the point $z=t$:
\eqn\mztexp{m^2_z(z,t)\sim{1\over
(z-t)^2}-2{(\Omega_z(z)-\omega_z(z))\over z-t}+{\rm zero}\enskip{\rm
modes}}
The zero modes in eq. \mztexp\ are multivalued, due to the fact that
$m_z(z,t)$ is multivalued on the Riemann surface.
Looking at eq. \mztexp, we try the following ansatz for $m^2_z(z,t)$:
$$m_z^2(z,t)=\partial^2_zK(z,t)-2\partial_zK(z,t)(\Omega_z(z)-
\omega_z(z)) +\psi_{zz}(z,t)+$$
\eqn\mztfin{\left(
\sum\limits_{i,j=1}^g\omega_i(z)\left|{\rm
Im}\enskip\Omega_{ij}\right|^{-1}\int_{\bar z_0}^{\bar z}\bar
\omega_j(\bar s)d\bar s\right)^2}
With this ansatz, at least the pole structure of $m_z^2(z,t)$ is
reconstructed. In fact,
if we expand the right hand side of eq. \mztfin\ at
$z=t$, the result is in agreement with eq. \mztexp. It remains a
zero mode contribution $\psi_{zz}(z,t)$, which satisfies the equation
$\partial_{\bar z}\psi_{zz}(z,t)=0$ and cannot be determined.
We notice also that in eq. \mztfin\ the quantity $\partial^2_zK(z,t)$ is
not a true tensor, because we do not have used the covariant derivatives
$\nabla_z$. A true tensor with a double pole in $z=t$ is
$\nabla_z^2K(z,t)$, but we do not need it, because the
expression of $m_z^2(z,t)$ contained in \mztfin\ should be inserted in
eq. \newexkzt\ and, when integrated, it yields:
$$\int_M d^2tg_{t\bar t}\nabla_z^2K(z,t)=
 \int_M d^2tg_{t\bar t}\partial_z^2K(z,t)=0$$
Using the above identity together with eqs. \scalthree\ and \scalfour,
we arrive at the following result:
$$\int_M d^2t g_{t\bar t}m^2_z(z,t)=\int_Md^2tg_{t\bar t}
\psi_{zz}(z,t)+A
\left(
\sum\limits_{i,j=1}^g\omega_i(z)\left|{\rm
Im}\enskip\Omega_{ij}\right|^{-1}\int_{\bar z_0}^{\bar z}\bar
\omega_j(\bar s)d\bar s\right)^2$$
Substituting this equation in \newexkzt\ we get:
\eqn\respartfin{\int_M d^2tg_{t\bar
t}(\partial_zK(z,t))^2=\int_Md^2tg_{t\bar t}\psi_{zz}(z,t)-
A\left(
\sum\limits_{i,j=1}^g\omega_i(z)\left|{\rm
Im}\enskip\Omega_{ij}\right|^{-1}\int_{\bar z_0}^{\bar z}\bar
\omega_j(\bar s)d\bar s\right)^2}
As a consequence of the fact that the left hand side of this equation is
singlevalued on the Riemann surface $M$, also the zero mode
$\psi_{zz}(z,t)$ should be singlevalued.
Therefore, it must be a linear combination of the $3g-3$ solutions
$\psi_{s,zz}(z,t)$ of the equation $\partial_{\bar z}\psi_{s,zz}(z)=0$
that are allowed on $M$.
This linear combination is of the following kind:
$$\psi_{zz}(z,t)=\sum\limits_{s=1}^{3g-3}f_s(t)\psi_{s,zz}(z)$$
where the functions $f_s(t)$, which depend only on the variable $t$, are
until now undetermined.
Inserting the above expression in eq. \respartfin, we have however still
a little simplification:
\eqn\resfin{\int_Md^2tg_{t\bar
t}(\partial_zK(z,t))^2=\sum\limits_{s=1}^{3g-3}\psi_{s,zz}(z)c_s}
with $c_s=\int_Md^2tg_{t\bar t}f_s(t)$.
In this way, remembering eq. \gzzm,
we have shown that on a Riemann surface the asymptotic form
at very short distances of the potential $G(z,w)$ governing the behavior
of the electrons is completely determined by the zero modes of the kind
$\psi_{s,zz}(z)$.
Unfortunately, we could not derive
the coefficients $c_s$ and, even using
the more sophisticated methods of ref. \ks, it seems not possible to
obtain their explicit form.
However, general physical considerations would suggest that $c_s=0$ for
$s=1,\ldots,3g-3$. In fact, zero modes are not expected to contribute to
observable effects.
\vskip 1cm
\newsec{CONCLUSIONS}
\vskip 1cm
In this paper we have quantized the abelian gauge field theories on a Riemann
surface for any nontrivial line bundle $P(M,U(1))$.
Despite of the fact that we applied the formalism developed here only to
the Schwinger model, the method is valid for any gauge field theory with
$U(1)$ gauge group of symmetry.
The only restriction is that the explicit form of the propagators of the
matter fields should be known.
For example, this is not the case of the massive fermions of scalar
fields.
Another technical difficulty is that perturbation theory on a manifold
is intrinsecally more complicated that in the flat case. As a matter of
fact, the flat connections $A^{\rm har}_\alpha$ should be treated as
external fields. They generate in this way new Feynamn graphs and there
is the problem of integrating over the moduli space of flat connections
in the path integral.
This difficulty is not present if we consider the Schwinger model on a
nontrivial line bundle. If $A_\alpha^{\rm I}=0$, in fact, the current
$J_\alpha^m$ of eq. \currents\ is conserved: $\partial_z J_{\bar
z}^m+\partial_{\bar z} J_z^m=0$.
Hence, $J_\alpha^m$ is a purely transversal vector and, using the
orthogonality properties of the Hodge decomposition, we have in eq.
\zqedfin:
$$\int_Md^2zA_\alpha^{\rm har}J^{m,\alpha}=0$$
As a consequence, the integrand in eq. \zqedfin\ is independent of
$\theta_i$ and $\phi_i$, so that the integration in these variables can
be factored out yielding the following generating functional:
$$Z_{\rm QED}[\bar\xi,\xi,k=0]=\int D\bar\psi
D\psi
{\rm
exp}\left\{-\left[\int_md^2z\left[\bar\psi_\theta(\partial_{\bar
z}+A^I_{\bar z})\psi_\theta+
\bar\psi^\theta\xi_{\theta}+\psi^\theta\bar\xi_\theta
\right]+\right.\right.$$
\eqn\zqedaiz{\left.\left.
{1\over 4}\int_Md^2zd^2z'\left(\bar\psi_{\theta}\psi_{\theta}G_{\bar
z'\bar z}(z,z')\bar\psi_{\bar \theta}\psi_{\bar\theta}\right)+{\rm
c.c}\right]\right\}}
Let us remember that eq. \zqedfin\ can be extended also to the case in
which the spinor fields are replaced by more general $b-c$ systems of
integer or half-integer conformal weights.
In this case, however, the matter fields have zero modes on a Riemann
surface and eq. \zqedfin\ cannot be simplified to eq. \zqedaiz\ even
on a trivial line bundle.
\smallskip
Another improvement with  respect to ref. \ferqed\
is that here we succeded in integrating over
the nonphysical gauge degrees of freedom in the path integral of the
Schwinger model. In this way, we have obtained an effective field theory
describing the dynamics of the two dimensional electrons on a Riemann
surface.
There are many interesting aspects of this theory which should be better
understood.
For example, the effective action:
$$S_{eff}=\int_Md^2z
\left[
\bar\psi_\theta(\partial_{\bar z}
+A{\bar z}
)\psi_\theta+
\bar\psi^\theta\xi_{\theta}
+\psi^\theta\bar\xi_\theta
\right]
+\int_Md^2zd^2z'\bar\psi_{\theta}
\psi_{\theta}
G_{\bar z'\bar z}(z,z')
\bar\psi_{\bar \theta}\psi_{\bar\theta}+
{\rm c.c}$$
where $A_{\bar z}=A^{\rm I}_{\bar z}+A^{\rm har}_{\bar z}$, should
describe an integrable model, since the Schwinger model is integrable.
As a matter of fact, the above action can be interpreted as the action
of a generalized version of the Thirring model.
Nevertheless, theintegrability of $S_{eff}$ on a Riemann surface
is not clear a priori.
Moreover, we still do not know the explicit form of the zero modes in
the fermionic sector when $c_1\ne 0$.
Since the main subject of this paper is the quantization of the gauge
field theories, we will answer these questions elsewhere.
Nonetheless, interesting physical informations concerning the high energy
behavior of the electrons on a Riemann surface have
already been extracted from
eqs. \sdbdisk, \sdbstwo\ and \resfin.
Unfortunately, we were not able to estimate the important zero mode
contribution appearing
in the asymptotic expression of the potential $G_{zw}(z,w)$ at
short distances
of eq. \resfin. However, we believe that it is possible to obtain its
explicit form representing the Riemann surface as an algebraic curve and
exploiting the methods developed in refs. \ref\ferstr{F. Ferrari, {\it
Int. Jour. Mod. Phys.} {\bf A5} (1990), 2799; {\it ibid.} {\bf A7}
(1992), 5131; {\it Comm. Math. Phys.} {\bf 156} (1993), 179.}.
\vskip 1cm
\newsec{ACKNOWLEDGEMENTS}
\vskip 1cm
The final version of this paper has been greatly benefitted by fruitful
conversations with C. M. Hull, M. Mintchev, O. A. Solov'ev and A. Wipf.
In particular I am grateful to C. M. Hull who suggested the use of the
Feynman gauge and O. A. Solov'ev, who drawed my attention to ref. \ks.
I am also grateful to M. B. Green for giving me the possibility of
talking at Queen Mary college on the subject of ${QED}_2$ on curved
space$-$times.
Finally, I would like to thank S. Theisen and J. Wess for their kind
hospitality at the Ludwig Maximilian University of Munich and for their
interest in my work.
This work was supported by a grant of Consiglio Nazionale delle
Ricerche, P.le A. Moro 7, Italy.
\listrefs
\bye

\centerline{REFERENCES}
\vskip 1cm
\item{[1]}

\item{[2]} M. F. Atiyah, R. Bott, {\it Phyl. Transl. R. Soc. Lond.}
{\bf A308} (1982), 523.\medskip
\item{[4]} M. Blau, G. Thompson, {\it Quantum Maxwell Theory on Arbitrary
Surfaces}, Preprint NIKHEF-H/91-08, MZ-TH/91-16.\medskip
\item{[5]} J. Soda, {\it Phys. Lett.} {\bf 267B} (1991), 214.\medskip
\item{[6]} E. Witten, {\it Two Dimensional Gauge Theories Revisited},
Preprint IASSNS-HEP-92/15.\medskip
\item{[9]} S. Pokorski, {\it Gauge Field Theories}, Cambridge University
Press, 1987.\medskip
\item{[11]} J. F. Cari\~nena, C. L\'opez, {\it Int. Jour. Mod. Phys.}
{\bf A7} (1992) 2447.\medskip
\item{[12]} L. Alvarez-Gaum\'e, G. Moore, C. Vafa, {\it Comm. Math. Phys.}
{\bf 106} (1986),1.\medskip
\item{[13]} E. Witten, {\it Comm. Math. Phys.} {\bf 121} (1989), 351.
\medskip
\item{[14]} M. Bonini, R. Iengo, {\it Int. Jour. Mod. Phys.}, {\bf A3} (1988),
841.\medskip
\item{[15]} E. Verlinde, H. Verlinde, {\it Nucl. Phys.} {\bf B288} (1987),
357.\medskip
\item{[17]} M. S. Chanowitz, {\it Phys. Lett.} {\bf 171B} (1986), 280.
\medskip
\item{[18]} J. Barcelos-Neto, A. Das, {\it Zeit. Phys. C} {\bf 32}
(1986), 527;
H. Leutwyler, {\it Phys. Lett.} {\bf 153B} (1985), 65;
F. M. Saradzhev, {\it Phys. Lett.} {\bf 278B} (1992), 449.\medskip
\item{[19]} A. H. Chamseddine, J. Fr\"ohlich, {\it Two Dimensional
Lorentz-Weyl Anomaly and Gravitational Chern-Simons Theory}, Preprint
ETH-TH/91-48, ZU-TH-30/91; A. M. Po\-lya\-kov, {\it Mod. Phys. Lett.} {\bf A2}
(1987), 899; V. G. Knizhnik, A. M. Polyakov, A. A. Zamolodchikov,
{\it Mod. Phys. Lett.} {\bf A3} (1988), 819;
F. David, {\it Mod. Phys. Lett.} {\bf A3} (1988), 1651;
J. Distler, H. Kawai, {\it Phys. Lett.} {\bf 221B} (1989), 509;
A. H. Chamseddine, {\it Phys. Lett.} {\bf 256} (1991), 379.\medskip
\item{[20]} J. M. F. Labastida, A. V. Ramallo, {\it Chiral Bosons Coupled
to Abelian Gauge Fields}, Preprint CERN-TH-5288/89.\medskip
\item{[21]} G. W. Bluman, R.D. Gregory, {\it Mathematika},
{\bf 32} (1985), 118.\medskip

\bye